\documentclass{kluwer}    
\usepackage[dvips]{graphicx}
\newdisplay{guess}{Conjecture}

\begin{document}
\begin{article}
\begin{opening}
\title{GAIA and Population II Visual Binaries}
\author{Antanas \surname{Bartkevi\v cius}}
\institute{Institute of Theoretical Physics and Astronomy, Go\v stauto 12, Vilnius 2009, Lithuania,
E-mail: bart@itpa.lt
\newline
Department of Theoretical Physics, Vilnius Pedagogical University, Student\c u 39, Vilnius 2340,
Lithuania}
\author{Art\=uras \surname{Gudas}}
\institute{Institute of Theoretical Physics and Astronomy,
Go\v stauto 12, Vilnius 2009, Lithuania, E-mail: arturas@itpa.lt}
\runningauthor{Antanas Bartkevi\v cius, Art\=uras Gudas}
\runningtitle{GAIA and Population II Visual Binaries}

\begin{abstract}
The results of a kinematical population classification of 497 Hipparcos orbital binaries are
discussed. Less than 10\% of the stars from the sample belong to the thick disk. Only seven stars
have halo kinematics. There are no direct dynamical mass determinations for extreme halo stars.
Some suggestions, concerning Population II visual binaries for which ground-based astrometric
observations in combination with GAIA data will be useful in the future for the better
determination of orbits for long period Population II binaries, are offered.
\end{abstract}
\end{opening}

\section{Introduction}
The ESA's GAIA mission will radically change our knowledge about visual binary stars and precise
stellar masses. It is expected that over 60 million visual binaries will be detected, 10 million
orbital elements calculated, and over 10000 star masses with uncertainties up to 1\% determined
\cite{GAIA}.

This contribution is aimed to show the unsatisfactory situation in the domain of the investigation
of Population II orbital binaries.  It is well-known that ground-based astrometric, radial velocity
and photometric observations in combination with Hipparcos data are very useful for the
considerable improvement of orbital, astrometric and photometric parameters of the known binaries
and the detection of new ones.  Similarly, new ground based astrometric and other observations of
the Population II long period visual binaries in combination with GAIA data will be useful in the
future for determining accurate orbits of these stars and their masses.  Several binaries are
proposed for observations.

\section{High Velocity Hipparcos Orbital Visual Binaries}
\inlinecite{BartkeviciusandGudas1} obtained kinematical parameters for \enlargethispage{2ex} a
sample consisting of 497 orbital visual binaries: stars from Hipparcos Catalogue Double Star Annex
DMSA/O \cite{HIP} and binaries with orbital elements obtained from reprocessed Hipparcos data and
ground-based observations (\opencite{Soderhjelm}; \opencite{Pourbaix};
\opencite{PourbaixandJorissen}; \opencite{MartinandMignard}; \opencite{Martinetal};
\opencite{Gontcharovetal}).  Kinematically, most (89\%) of the selected binaries belong to the thin
disk, less than 10\% to the thick disk. Only seven binaries are halo stars. Five of them
kinematically belong to the intermediate halo. Two sdF7 subdwarfs HD 219617 and HD 132475 have
extreme halo kinematics. Unfortunately, these very interesting stars have rather  poorly determined
orbital elements. HD 219617's orbit inclination is $i=90^{\circ}$ \cite{Hartkopfetal},
\inlinecite{Heintz} presented two alternative orbits, $P=260$ yr, $a=0.800''$, and $P=100$ yr,
$a=0.402''$. For the second subdwarf with extreme halo kinematics HD 132475, even its orbit is not
confirmed by subsequent observations \cite{WorleyandHeintz}. The search through the catalogues of
metal-deficient stars (\opencite{Bartkevicius1}, \citeyear{Bartkevicius2}), Population II Star
Catalog \cite{BartkeviciusandBartkeviciene}, unpublished Population II Binary Catalog
\cite{Bartkevicius3} and the papers published after compilation of this catalog do not reveal any
more extreme Population II visual binaries with determined orbits. Therefore, we completely lack
direct dynamical mass determination for these astrophysically very important stars.

As mentioned above, the orbital elements of binaries analysed by \inlinecite{BartkeviciusandGudas1}
have been determined from Hipparcos measurements or combining them with the ground-based orbits.
These binaries form only one third of the orbits published in a new Fifth Catalogue of Orbits of
Visual Binary Stars \cite{Hartkopfetal}.  In spite of the fact that remaining two-thirds of the
orbits have been obtained only from the ground-based measurements and in most cases are less
precise than the orbits of stars analysed by \inlinecite{BartkeviciusandGudas1}, kinematical
classification of those stars will be performed \cite{BartkeviciusandGudas2} with the hope to
supplement the sample of the thick disk orbital binaries. \nopagebreak[4]

\section{Metal-deficient Visual Binaries, Possible Candidates for Ground-based
Observations}

GAIA measurements will allow us to obtain reliable orbits for visual binaries with orbital periods
shorter than 50 years. Just as in the Hipparcos case, the ground-based astrometric measurements in
combination with GAIA measurements will significantly improve the precision of orbital elements,
especially for long period binaries. These measurements are very important for the rare Population
II binaries.

Eleven metal-deficient visual binaries candidates for the ground-based measurements have been
selected from the unpublished Population II Binary Catalogue \cite{Bartkevicius3}. The criteria of
the selection of stars were as follows: $[Fe/H]<-0.6$, no orbit determinations, number of existing
position measurements $>5$, and being a physical binary. Relative astrometric measurements
(position angle and angular separations of the components) have been obtained from the Washington
Double Star Data Center.  The mean angular separation $<D>$ (in arcsec) and the rate of change of
position angle $dPA/dt$ (deg/yr) have been calculated. A rough estimate of orbital period has been
made from the rate of change of position angle $dPA/dt$ and from the third Kepler's law, assuming
the mean projected separation $<D>$ as major semi-axis of the orbit, mass sum that equals to 1 Sun
mass, and using trigonometric parallax from Hipparcos.  Calculated elements with spectral type,
metallicity $[Fe/H]$, number of angular separation $ND$ and position angle $NPA$ measurements are
presented in Table 1.

\begin{table*}[h]
{\scriptsize\selectfont
\caption[]{Metal-deficient visual binary stars -- candidates for the ground-based observations.}
\label{sphericcase}
\begin{tabular}{r@{\extracolsep{2pt}}r@{\extracolsep{0pt}}l@{\extracolsep{2pt}}l@{\extracolsep{5pt}}r@{\extracolsep{0pt}}l@{\extracolsep{2pt}}r@{\extracolsep{0pt}}l@{\extracolsep{2pt}}r@{\extracolsep{3pt}}r@{\extracolsep{1pt}}l@{\extracolsep{2pt}}r@{\extracolsep{4pt}}r@{\extracolsep{1pt}}l}
\hline
  &Name    &    &  Sp     &  \multicolumn{2}{c}{$[Fe/H]$}&  \multicolumn{2}{c}{$<D>$}  & ND & \multicolumn{2}{c}{$dPA/dt$} & NPA  & \multicolumn{2}{c}{$P$}\\
  &        &    &         &    &   & \multicolumn{2}{c}{arcsec} & & \multicolumn{2}{c}{deg/yr}& &\multicolumn{2}{c}{yr}\\
\hline
HD&    23439&AB  &sdK1+sdK3&  $-$1&.12&     7&.94  &    95&   0&.06  &   95&    3000&        \\
HD&   111971&AB  &MDE      &  $-$0&.82&     1&.38  &    16&  $-$0&.03  &   16&    1000&        \\
HD&   111980&AB  &sdF7     &  $-$1&.25&     1&.95  &     7&  $-$0&.20: &    7&    1900&        \\
HD&   128636&AB  &F-WMM    &  $-$0&.86&     0&.29  &     6&    &     &    6&     100&        \\
HD&   130166&AB  &G2WF5    &    &   &     0&.49  &     5&    &     &    5&     200&        \\
HD&   156384&AB-C&M2       &    &   &    31&.01  &     9&   0&.08  &   10&    3500&        \\
HD&   156384&AC  &M2       &  $-$0&.5 &    31&.16  &    47&   0&.09  &   47&    3500&        \\
HD&   156384&BC  &         &    &   &    31&.3   &    15&   0&.14  &   14&    3000&        \\
HD&   163810&AB  &sdG5     &  $-$1&.47&     0&.44  &    12&  $-$1&.14  &   12&     300&        \\
HD&   211998&AB  &sdG2     &  $-$1&.50&     0&.12  &    11& $-$72&::  &    5&       5&        \\
HD&   224927&AB  &sdA8     &  $-$1&.08&     0&.15  &     6&  $-$3&.:   &    4&     200&        \\
 G& 19-20/21&    &wdA+sdM6 &    &   &    12&.92  &     7&    &     &    7&   10000&        \\
 G&    122-2&AB  &(sdK1)   &    &   &     3&.76  &    32&   0&.13  &   32&    2000&        \\
\hline
\end{tabular}
}
\end{table*}

The most promising stars candidates with estimated period less than 1000 years are HD 128636, HD
130166, HD 163810 and HD 224927. Very interesting candidate system is sdG5 subdwarf HD 163810. Its
estimated period is about 300 years. An exceptional system is $\nu$ Ind (HD 211998) having a short
period of about 5 years.

\section{Conclusions}

1. Less than 10\% of Hipparcos orbital visual binaries kinematically belong to the thick disk. Only
7 of 497 Hipparcos orbital visual binaries have halo kinematics. Direct dynamical mass
determinations for extreme halo stars still do not exist.

2. In order to take a full advantage of GAIA measurements and enlarge the number of reliably
derived long period halo binary orbits, relative ground-based astrometric measurements should be
made. Eleven candidate systems are proposed in Table 1.

\begin{acknowledgements}
We are indebted to V. Vansevi\v cius for his valuable comments. This research has made use of the
Washington Double Star Catalog maintained at the U.S. Naval Observatory.
\end{acknowledgements}

\pagebreak
\end{article}
\end{document}